\documentclass[prb,preprint]{revtex4-1} 

\usepackage{amsmath}  
\usepackage{amsfonts} 
\usepackage{graphicx} 
\usepackage{subcaption}
\usepackage{url}

\newcommand{\ie}{{\it i.e.,\ }}
\newcommand{\eg}{{\it e.g.,\ }}

\begin{document}


\title{Experimental determination of circuit equations}

\author{Jason Shulman}
\email{jason.shulman@stockton.edu} 
\altaffiliation[permanent address: ]{101 Vera King Farris Dr., 
  Galloway, NJ 08205} 
 
\author{Frank Malatino} 

\author{Matthew Widjaja}

\affiliation{Department of Physics, Richard Stockton College of New Jersey, Galloway, NJ 08205}

\author{Gemunu H. Gunaratne}

\affiliation{Department of Physics, University of Houston, Houston, TX 77204}


\date{\today}

\begin{abstract}
Kirchhoff's laws offer a general, straightforward approach to circuit analysis. Unfortunately, use of the laws becomes impractical for all but the simplest of circuits. This work presents a novel method of analyzing direct current resistor circuits. It is based on an approach developed to model complex networks, making it appropriate for use on large, complicated circuits. It is unique in that it is not an analytic method. It is based on experiment, yet the approach produces the same circuit equations obtained by more traditional means.
\end{abstract}

\maketitle 

\section{Introduction} 

In introductory physics, students are universally taught to analyze circuits using Kirchhoff's laws. The standard approach employs the loop and current laws in order to derive a set of equations that can be solved for the branch currents. This technique is, of course, very powerful in that it provides a systematic procedure for studying a variety of circuits. Unfortunately, it becomes unfeasible for any circuit of modest complexity. We are forced to search for an alternative method and/or engage a computer. There have been several interesting and clever alternatives presented in the literature that aim to teach or supplement the standard methodology presented in textbooks.\cite{AJPrandomwalk,AJPnode,AJPvariational,BookerText} Here, we take a more bottom-line approach and present, not an analytical method, but an experimental one to obtain the equations that govern circuit behavior. These are the same equations found through an implementation of Kirchhoff's laws. The experimental method is based on  a modeling approach developed to aid in controlling large, complex networks.\cite{GunaratnePLOS} Thus, it is well suited for quite complicated circuits. It, however, is not intended to replace the learning and use of standard techniques; it simply provides an efficient means of determining the equations when other methods of analysis are not practical or even feasible. This practicality is also accompanied by several didactic aspects, allowing students to gain insight into the nature of circuits and the method. 

The network modeling methodology is presented in section \ref{NetworkSection}. Section \ref{CircuitSection} reviews the node method, a more traditional approach to circuit analysis. The results of both the circuit and network methods are compared in section \ref{CompareSection}. Finally, section \ref{ResultsDiscussion} demonstrates the implementation of the network approach on a simple circuit.

\section{A network model}
\label{NetworkSection}
\subsection{General considerations}
\label{ModelSection}
The network modeling approach, upon which the experimental method is based, was developed specifically for genetic networks,\cite{GunaratnePLOS,ShulmanBiophysJ, ShulmanBiophysRev} where accurate models are difficult or impossible to construct from experimental data.\cite{Gardner} We have since shown that the approach is appropriate for a host of large, nonlinear networks.\cite{unpub} Circuits can certainly fall into this category; however, in the present work, we will restrict our attention to those appropriate for undergraduate students beyond the introductory courses; that is, we will focus on networks of linear (ohmic) resistors in dc circuits. First, we will briefly outline the network modeling methodology.

\begin{figure}[h!]

        \centering
        \begin{subfigure}[b]{0.5\textwidth}
                \centering
                \includegraphics[width=\textwidth]{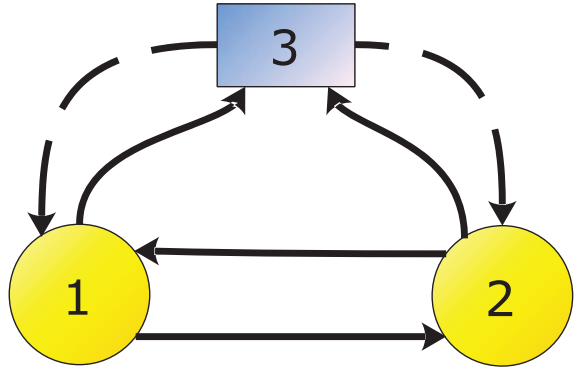}
                \caption{}
                \label{ExampleNet:net}
        \end{subfigure}%
        \\ 
        \begin{subfigure}[b]{0.5\textwidth}
                \centering
                \includegraphics[width=\textwidth]{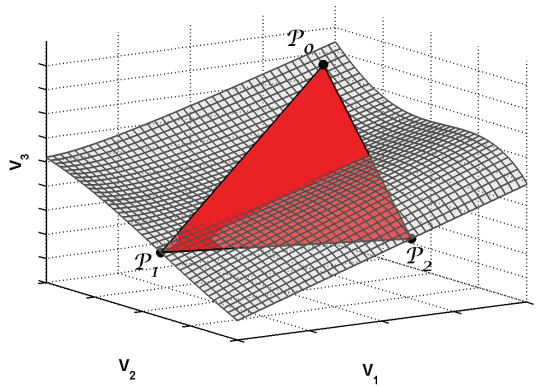}
                \caption{}
                \label{ExampleNet:plot}
        \end{subfigure}
      \caption{(Color online) (\subref{ExampleNet:net}) A simple three node network. In the network model, node three (the slave) is exclusively controlled by nodes one and two (the master nodes). Solid arrows represent interactions approximated by the network model. Dashed arrows, from the slave, are ignored. (\subref{ExampleNet:plot}) The value $V_3$ of the slave node is determined by those of the master nodes. A surface, representing the values of the slave parameter, is formed upon sweeping the master node values. A plane, approximating the surface, can be determined by a few experiments. The plane is a good approximation to the surface if the surface is sufficiently smooth.}
\end{figure}

The details of the network modeling approach are provided elsewhere.\cite{GunaratnePLOS,ShulmanBiophysJ, ShulmanBiophysRev} Here, we present only that which is relevant to our purpose. Consider the three node network in Fig. \ref{ExampleNet:net}. A parameter $V$ is associated with each node. Later, when discussing circuits specifically, $V$ will represent the potential at the node, and each arrow will symbolize an electrical component connecting two nodes. The vector $\mathbf{V}=\left[V_1, V_2, V_3 \right]$ defines the state of the system. Each arrow represents the interaction between two nodes, corresponding to a possibly unknown nonlinear mathematical function. Changing $V_1$, for example, will induce a change in the values of $V_2$ and $V_3$, as determined by these functions. While the set of interactions and the associated mathematical functions are commonly unknown to us, we can often perform a set of experiments on the system in which some of the node parameters $V$ are perturbed, and the corresponding effect on the remaining parameters is measured. An example of this would be fixing the potential at one point in a circuit and measuring the potential at the others in response to this change. When choosing nodes to perturb, we often select highly connected hub-like nodes which are able to influence many others. Imagine such an experiment on the three node network. Hub nodes cannot exist in such a small system; however, we choose to perturb nodes one and two (the circles) and observe the effect on node three (the square). Upon sweeping the values $V_1$ and $V_2$, a two dimensional surface can be formed (Fig. \ref{ExampleNet:plot}). When presented in this way, $V_1$ and $V_2$ are the independent variables, and $V_3$ is the dependent variable. In other words, the value $V_3$ is determined by $V_1$ and $V_2$. The independent nodes are referred to as the master nodes (circles), which determine the behavior of the remaining slave nodes (square). In our network model, we make the approximation that the values of the slave nodes are exclusively determined by those of the master nodes. This approximation corresponds to ignoring the dashed lines in Fig. \ref{ExampleNet:net}, which run from the slave nodes to the masters. In keeping with the approximation, any slave/slave interactions will be ignored as well. The categorization of nodes into masters and slaves is not limited to small networks such as the one described here. Networks with an arbitrary number of master/slaves can be considered. In such cases, the surfaces corresponding to Fig. \ref{ExampleNet:plot} will be higher dimensional. By determining surfaces in this way, one can construct a model of the (directed) interactions between the highly coupled master nodes and the sparsely coupled slaves.

Unfortunately, it is often not feasible to perform the experiment imagined above for large systems. Such experiments on gene networks, for example, would be prohibitively expensive.\cite{Gardner, Gardner2} However, these surfaces exist in principle, and one can generally find an approximation to the surface by conducting only a few experiments. Consider again the network of Fig. \ref{ExampleNet:net}. Let the state of the system be $\mathbf{V^{(o)}}=\left[V^{(o)}_1, V^{(o)}_2, V^{(o)}_3 \right]$ when the unperturbed network is measured. This state corresponds to one point ($\mathcal{P}_o$) on the surface in Fig. \ref{ExampleNet:plot}; the point at which the values of nodes one and two equal their unperturbed values. It is often relatively easy to perturb a system by fixing the parameter of one node to zero, \eg grounding a node in a circuit. By forcing $V_1 = 0$, the remaining parameters will be modified from their natural values, and the state of the network will be $\mathbf{V^{(1)}}=\left[0, V^{(1)}_2, V^{(1)}_3 \right]$, where the superscript indicates the modification of node 1. This corresponds to point $\mathcal{P}_1$ on the surface. Similarly, node two can be fixed to zero to obtain the state $\mathbf{V^{(2)}}$ and point $\mathcal{P}_2$. These three points, lying in a three dimensional space, uniquely define a plane. Furthermore, the points lie on both the plane and the surface, indicating close proximity between the two in the regions surrounding the points. The plane represents a good approximation to the surface if the surface is sufficiently smooth. While one may not be able to determine the surface, the plane can be found conducting relatively few experiments, in this case by measuring the state of the system three times, when it is unperturbed and upon perturbation of each master node. A series of linear equations can be written which describe the plane and relating the values of the slave node parameters to those of the master nodes. For the present case, in which there is only one slave node, a single equation approximating the dependence of $V_3$ on $V_1$ and $V_2$ is produced, 
\begin{equation}
\label{SlaveEx}
V_3 - V^{(o)}_3 = B_{11}\left(V_1 - V^{(o)}_1\right) + B_{12}\left(V_2 - V^{(o)}_2\right).
\end{equation}
Here, the coefficients $\mathbf{B}$ describe the plane, and we have subtracted the unperturbed values $\mathbf{V^{(o)}}$, essentially stating that the deviation of $V_3$ from its unperturbed value is dependent on the deviations of $V_1$ and $V_2$ from theirs. Eq. (\ref{SlaveEx}) relates the slave parameters to those of the master nodes. For the model to be complete, one needs to account for interactions between the master nodes. In keeping with the linear approximation, we assume relationships between the master nodes take the form,
\begin{subequations}
\label{MasterEx}
\begin{eqnarray}
\label{MasterEx:a}
\dot{V}_1 & = A_{11}\left(V_1 - V^{(o)}_1\right) + A_{12}\left(V_2 - V^{(o)}_2\right)\\
\label{MasterEx:b}
\dot{V}_2 & = A_{21}\left(V_1 - V^{(o)}_1\right) + A_{22}\left(V_2 - V^{(o)}_2\right),
\end{eqnarray}
\end{subequations}
where the coefficients $\mathbf{A}$ represent effective interactions between the master nodes. 

This procedure generalizes nicely to larger systems. By measuring the states of the unperturbed network and situations in which each of the $n$ master nodes is singly modified, we obtain $n+1$ points which lie on the surface and form an \textit{n}-dimensional plane or, equivalently, the elements of the $\mathbf{A}$ and $\mathbf{B}$ matrices. This plane is the geometric representation of the network model, which describes the response of the system to modification of the master nodes. The plane is not the surface; however, it represents a reasonable approximation if the surface is relatively smooth. Of course, this approximation can be improved, by including quadratic (or higher) terms at the cost of performing more experiments to determine the new parameters.

\subsection{Exact models}
\label{ExactModelSection}
The approach described above does not generally produce an exact model of the network, rather an approximation to the system is generated, the quality of which is quantified by the proximity of the plane and the surface. The methodology has been shown to accurately represent several types of networks.\cite{GunaratnePLOS,ShulmanBiophysJ, ShulmanBiophysRev} Furthermore, the models cease to be approximate under two conditions: 1) the surface itself is a plane and 2) all nodes are considered to be master nodes. The first condition is equivalent to having all interactions between nodes be linear in nature. In such a situation, the procedure described above fits a plane on to a planar surface. The second condition removes one of the key assumptions employed when constructing the model, namely that some set of nodes, the masters, exclusively control those considered to be slaves. The removal of this approximation, combined with a linear system, results in an exact mathematical representation of the network.\cite{NoteExact} As there are no slave nodes, equations similar to Eq. (\ref{SlaveEx}) are not present in the model, and the dynamics of the nodes is described by the master node equations (Eq. (\ref{MasterEx})),
\begin{equation}
\label{MasterEq}
\dot{V}_i = \sum_{j}A_{ij}\left(V_j - V^{(o)}_j\right),
\end{equation}
where the sum is over all nodes. Under steady state conditions, all $\dot{V}_i=0$, and Eqs. (\ref{MasterEq}) simplify to, 
\begin{equation}
\label{SSMasterEq}
0 = \sum_{j}A_{ij}\left(V_j - V^{(o)}_j\right).
\end{equation}
This represents an exact mathematical description of a linear system in the steady state. Thus, by performing the experiments described in section \ref{ModelSection}, the matrix $\mathbf{A}$  can be calculated (see Supplementary Materials\cite{SupplementaryMaterials}). The equations describing the behavior of the network will, therefore, have been determined. 

The equations can be put in a more convenient form by solving for the node parameters themselves. For example, in the case of a three node network,
\begin{subequations}
\label{MasterEqEx}
\begin{eqnarray}
V_1 = -\frac{A_{12}}{A_{11}}V_2 - \frac{A_{13}}{A_{11}}V_3 + C_1\\
V_2 = -\frac{A_{21}}{A_{22}}V_1 - \frac{A_{23}}{A_{22}}V_3 + C_2\\
V_3 = -\frac{A_{31}}{A_{33}}V_1 - \frac{A_{32}}{A_{33}}V_2 + C_3,
\end{eqnarray}
\end{subequations}
where the values $C_i$ are constants containing the elements of $\mathbf{A}$ and $\mathbf{V}^{(o)}$.

Here, we will demonstrate that this procedure can be used to determine the equations for a circuit of resistors, even when implementation of Kirchhoff's laws is not practical. For a circuit containing $N$ nodes, $N+1$ experiments in which the potential $V$ of each node is measured, need to be performed. It is not necessary to fix the node values to zero, as in the example described above. Any value can be used to construct the model; however, in the context of circuits, setting the node potential to zero (by grounding) is quite convenient and will be used throughout the work. The experiments described here can be performed efficiently with a computer and an analog/digital card.  If such resources are unavailable, two people using a handheld multimeter, \eg lab partners, can complete the experiments quickly as well. As a bonus, we find that applying these ideas to circuits provides some physical insight into the parameters of the model ($\mathbf{A}$), something which has been elusive previously.

\section{Circuits and the node method}
\label{CircuitSection}

\begin{figure}[h!]
\centering
\includegraphics[width=5in]{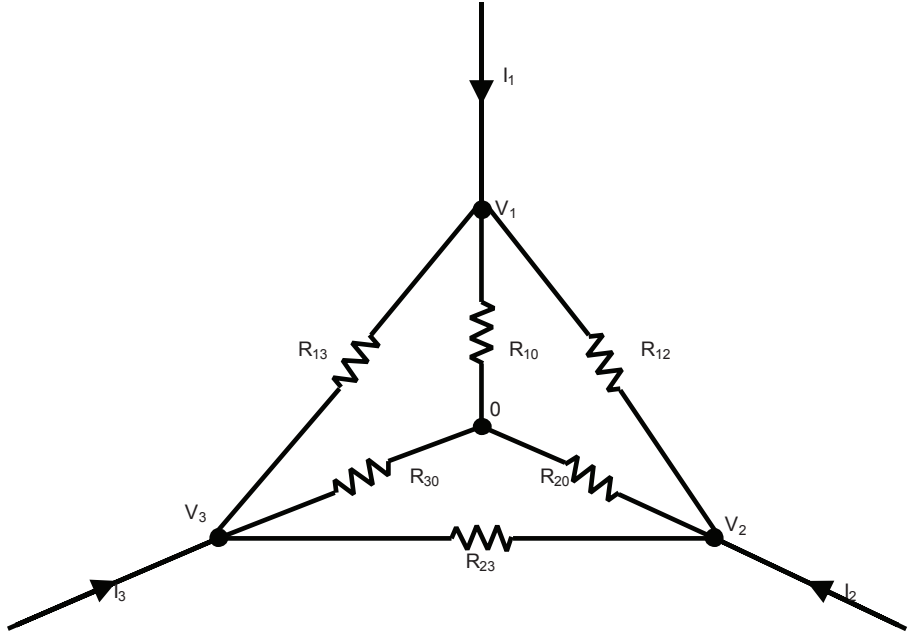}

\caption{Current is injected into nodes one, two, and three and exits via node zero, which is grounded. Potentials $V_1$, $V_2$, and $V_3$ are formed at the nodes.}
\label{GeneralCircuit}
\end{figure}

Before we begin applying these ideas to circuits, it will be instructive to examine some properties of electrical networks and their equations. Consider the four node network shown in Fig. \ref{GeneralCircuit}. Direct currents $I_1$, $I_2$, and $I_3$ flow into the network as shown and exit via node zero, which is held at zero potential. The potentials of the remaining nodes are $V_1$, $V_2$, and $V_3$. For simplicity, we will utilize the node method\cite{BookerText, FoundationsText}, a technique not typically taught in undergraduate courses, to analyze this network. The presentation will follow that of Ref. \onlinecite{BookerText}.

Kirchhoff's current law ensures that the current flowing into node one, $I_1$, equals the sum currents flowing through $R_{10}$, $R_{12}$, and $R_{13}$. This observation leads one to conclude
\begin{equation}
\label{initialI1}
I_1 = \frac{V_1-0}{R_{10}}+\frac{V_1-V_2}{R_{12}}+\frac{V_1-V_3}{R_{13}}.
\end{equation}
Similar considerations lead to equations for nodes two and three,
\begin{subequations}
\begin{eqnarray}
I_2 = \frac{V_2-0}{R_{20}}+\frac{V_2-V_1}{R_{12}}+\frac{V_2-V_3}{R_{23}}\\
I_3 = \frac{V_3-0}{R_{30}}+\frac{V_3-V_1}{R_{13}}+\frac{V_3-V_2}{R_{23}}.
\end{eqnarray}
\end{subequations}
Upon rearrangement, the incoming currents can be written, 
\begin{subequations}
\label{IRV}
\begin{eqnarray}
I_1 = \left(\frac{1}{R_{10}}+\frac{1}{R_{12}}+\frac{1}{R_{13}}\right)V_1 - \frac{1}{R_{12}}V_2 - \frac{1}{R_{13}}V_3\\
I_2 = - \frac{1}{R_{12}}V_1 + \left(\frac{1}{R_{20}}+\frac{1}{R_{12}}+\frac{1}{R_{23}}\right)V_2  - \frac{1}{R_{23}}V_3\\
I_3 = - \frac{1}{R_{13}}V_1 - \frac{1}{R_{23}}V_2 + \left(\frac{1}{R_{30}}+\frac{1}{R_{13}}+\frac{1}{R_{23}}\right)V_3.  
\end{eqnarray}
\end{subequations}
By replacing the resistances, $R_{ij}$ with the corresponding conductances, $G_{ij} = 1/R_{ij}$, the equations can be placed in a simplified form,
\begin{subequations}
\label{IGV}
\begin{eqnarray}
I_1 = G_{11}V_1 - G_{12}V_2 - G_{13}V_3\\
I_2 = - G_{12}V_1 + G_{22}V_2  - G_{23}V_3\\
I_3 = - G_{13}V_1 -G_{23}V_2 +G_{3}V_3,
\end{eqnarray}
\end{subequations}
where 
\begin{equation}
\label{SelfCond}
G_{11} = G_{10}+G_{12}+G_{13}
\end{equation}
and is called the coefficient of self-conductance for node one.\cite{NoteCoeff} The self-conductances of the other nodes are similar and can be obtained from the terms within the parentheses of their respective equations. It is worth noting that the coefficient of self-conductance for a node is the sum of conductances from all resistive channels connecting the node in question to the others in the network. \cite{NoteCoeff2} This fact will become important when examining the physical significance of the parameters in the network equations.

The situation represented in Fig. \ref{GeneralCircuit} is not typically realized in the laboratory or classroom.  More commonly, the network is connected to a single voltage source, and current enters through one or more nodes. It is then distributed throughout network and returns to the power supply. If, for example, only node one is connected to the power supply, held at potential $V_s$, via a resistor $R_{1s}$, the current $I_1$ is not fixed but depends on the potential difference across $R_{1s}$, \ie $\left(V_s - V_1\right)/R_{1s}$. The remaining currents are zero since no current is being injected into the other nodes. In this case, the updated equation for node one reads, $G_{1s}V_s = G_{11}V_1 - G_{12}V_2 - G_{13}V_3$, and, with the additional connection, the coefficient of self-conductance has another term, $G_{11} = G_{10}+G_{12}+G_{13} +G_{1s}$.

The equations (\ref{IGV}), updated to incorporate a power supply, can be solved for the node potentials to obtain,
\begin{subequations}
\label{CircuitEqs}
\begin{eqnarray}
\label{CircuitEqs:a}
V_1 = \frac{G_{12}}{G_{11}}V_{2} + \frac{G_{13}}{G_{11}}V_{3} + C^\prime_1\\
V_2 = \frac{G_{12}}{G_{22}}V_{1} + \frac{G_{23}}{G_{22}}V_{3} + C^\prime_2\\
V_3 = \frac{G_{13}}{G_{33}}V_{1} + \frac{G_{23}}{G_{33}}V_{3} + C^\prime_3,
\end{eqnarray}
\end{subequations}
where the $C^\prime_i$ are constants (possibly zero). In this form, the equations relate the potential at one node to the potentials of the others and a constant. They can be used in a variety of ways. For example, they can be solved simultaneously to obtain the node potentials of the unperturbed circuit. Further, if the circuit is perturbed by externally fixing the potentials of nodes two and three, Eq. (\ref{CircuitEqs:a}) will determine the new potential $V_1$.

\section{Network and circuit equations}
\label{CompareSection}

If one needs to analyze a complex circuit of resistors, it might not be convenient to employ Kirchhoff's laws. One can be assured, however, that equations in the form of Eqs. (\ref{CircuitEqs}) will describe the node potentials. Alternatively, the network strategy described in sections \ref{ModelSection} and \ref{ExactModelSection} could be implemented to obtain network equations which would also describe the circuit. Specifically, for the circuit in Fig. \ref{GeneralCircuit}, one would obtain Eqs. (\ref{MasterEqEx}). Note the similarity in the form of Eqs. (\ref{MasterEqEx}) and (\ref{CircuitEqs}). Both systems represent an exact description of the circuit and demonstrate the node potential is a linear combination of the other potentials. The coefficients preceding the node potentials as well as the constants in each system of equations must be equal, \ie
\begin{subequations}
\label{Coefs}
\begin{eqnarray}
\label{Coefs:a}
-\frac{A_{ij}}{A{ii}} = \frac{G_{ij}}{G{ii}}\\
\label{Coefs:b}
C_i = C^\prime_i.
\end{eqnarray}
\end{subequations}
These equations represent a major conclusion of this work. By performing the $N+1$ experiments described above and calculating the matrix $\mathbf{A}$, one can obtain the equations describing the circuits without Kirchhoff's laws. We are assured, through Eqs. (\ref{Coefs}), that these equations will be the same as those obtained through more conventional means.

It is important to recognize that Eqs. (\ref{Coefs}) represent the most fundamental relationship between the circuit and network equations. For example, one cannot equate $A_{ij}$ and $G_{ij}$. First, the $A_{ii}$ presented here are equal to one by convention,\cite{SupplementaryMaterials} which is not generally true for $G_{ii}$. Second, a glance at their respective equations indicates that their units are different, \ie $\mathrm{s}^{-1}$ for $A_{ij}$ and $\Omega^{-1}$ for $G_{ij}$. 

We will see below that this equivalence between the circuit and network equations can be exploited to easily obtain the equations governing circuit behavior. While this is the central focus of the work, it is also important to extract any physical insight, an important lesson for students. It has already been noted that $G_{ii}$ is the sum of all conducting channels connecting to node $i$. It is a measure quantifying the connectivity of node $i$ to the rest of the circuit. It says nothing, however, about how those connections are distributed throughout the circuit. For example, a node with a single connection could possess a coefficient of self-conductance with the same value as a node with five connections. The distribution information is contained in $G_{ij}/G_{ii}$, the coefficients of Eqs. (\ref{CircuitEqs}). Consider the circuit described by these equations. Written explicitly, the first coefficient of (\ref{CircuitEqs}a) is, $G_{12}/G_{11} = G_{12}/\left(G_{10}+G_{12}+\ldots+G_{1s}\right)$. In other words, $G_{12}$ is some portion of $G_{11}$, and, therefore, $G_{12}/G_{11}$ describes the strength of the connection between nodes one and two relative to all the other connections nodes one makes with the rest of the circuit. It is a measure of how much influence node two has on one compared to the other nodes. Note also, $\sum_jG_{ij}/G_{ii}=100\%$. This has all been gained by examining the constants in the circuit equations. In light of Eq. (\ref{Coefs:a}), $A_{ij}/A_{ii}$, the coefficients in the network equations, describe the proportion the $ij$ channel contributes to the total connectivity of node $i$. More generally, such a situation arises when the interaction between two nodes in a network is proportional to the \textit{difference} in the node parameters, as is the case with circuits. This physical insight into the constants of the network equations has been unavailable in prior studies.

Equation (\ref{Coefs:b}) indicates the constants in the network equations ($C_i$) represent the contributions from the power supply and ground, \ie $C_i = \left(G_{js}/G_{ii}\right)V_s$. For the example circuit described above, in which the power supply only connects to node one, only $C_1\neq0$ since the other nodes make no connection to the supply $\left(G_{2s}  = G_{3s} = 0\right)$.

\section{Results and discussion}
\label{ResultsDiscussion}
\subsection{A simple example}
\begin{figure}[h!]
\centering
\includegraphics[width=5in]{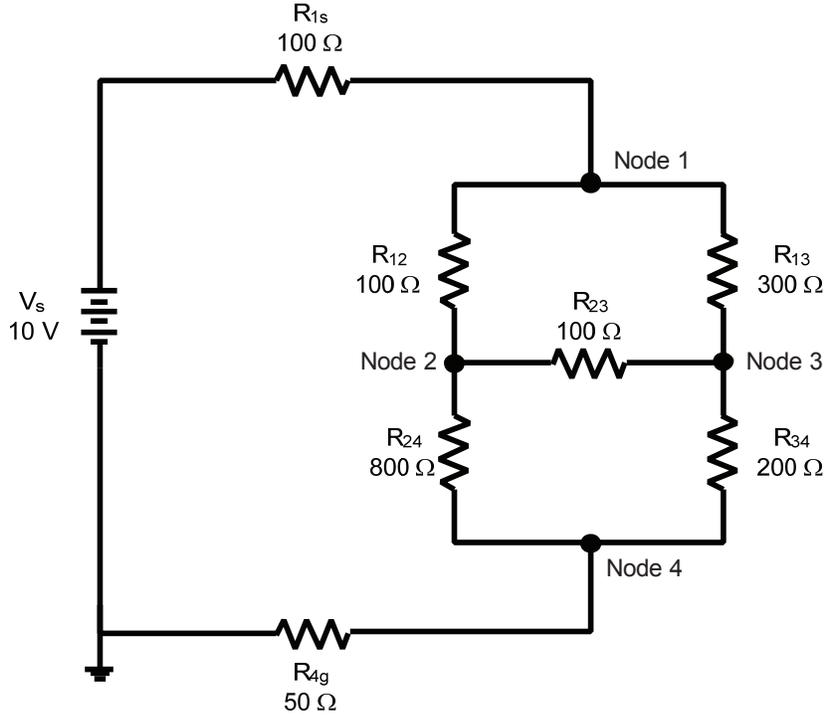}

\caption{The circuit used to generate data in Table \ref{ExampleData}.}
\label{ExampleCircuit}
\end{figure}

We will illustrate the ideas discussed above with the simple four node circuit shown in Fig. \ref{ExampleCircuit}. Such a circuit is simple enough to be studied with the standard techniques, and the results of such an analysis will be compared to that of the network method. In what follows, the results of simulations,\cite{Simulation} rather than experiments, will be presented so complicating factors such as experimental error and resistor tolerance will not hinder the exposition, hopefully allowing the reader to easily verify the calculations. For every simulation described, the corresponding experiment has been performed (by undergraduate students) and results verified (except for section \ref{BurnedResistorDetection}).  

A student in an introductory course would use Kirchhoff's laws to calculate the unknown branch currents and the potential differences across the resistors. A more advanced analysis could utilize the node method to obtain equations which illustrate the relationships, or interactions, between the nodes, \eg Eqs. (\ref{CircuitEqs}). For the circuit of Fig. \ref{ExampleCircuit}, the equations are,
\begin{subequations}
\label{ExampleCircuitEqs}
\begin{eqnarray}
V_1 = 0.43V_2+0.14V_3+4.29\\
V_2 = 0.47V_1+0.47V_3+0.06V_4\\
V_3 = 0.18V_1+0.55V_2+0.27V_4\\
V_4 = 0.05V_2+0.19V_3.
\end{eqnarray}
\end{subequations}

Alternatively, the network method could be implemented to determine the equations, which would take the form of Eqs. (\ref{SSMasterEq}). The primary task is to determine the elements of $\mathbf{A}$ and $\mathbf{V^{(o)}}$. This requires experimental input, which consists of the node potentials of the unperturbed circuit $\left(\mathbf{V}^{(o)}\right)$ and circuits in which each node potential has been individually fixed. Here, each node will be grounded, \ie $V=0~\mathrm{V}$. The results of these five experiments are contained in Table \ref{ExampleData}. These data can be used to calculate the elements of $\mathbf{A}$. We have included an algorithm and an explanation of the calculation in the Supplementary Materials.\cite{SupplementaryMaterials} For the circuit in Fig. \ref{ExampleCircuit}, 
\begin{equation}
\label{ExampleA}
\mathbf{A} =
\begin{pmatrix}
     1 & -0.43 & -0.14 & 0 \\
-0.47 & 1 & -0.47 & -0.06 \\
-0.18 & -0.55 & 1 & 0.27 \\
0 & -0.05 & -0.19 & 1 \\
\end{pmatrix}.
\end{equation}
Note the similarity of the elements to coefficients of Eqs. (\ref{ExampleCircuitEqs}). This is a consequence of the relationship between the circuit and network equations, Eq. (\ref{Coefs:a}). The network equations can be expanded and solved for the potentials to obtain the same equations found using the node method (Eqs. \ref{ExampleCircuitEqs}). Once these equations have been determined, by either method, all parameters of the circuit can be calculated. They can be solved for the node potentials of the unperturbed circuit $\left(\mathbf{V}^{(o)}\right)$ or any perturbed circuit in which one or more potentials are fixed. Branch currents can be calculated with the node potentials and the resistor values. This particular example is simple, and neither the network technique nor the node method is required to analyze the circuit. For more complicated circuits, however, the network method is quite straightforward and can be implemented quickly. 

\begin{table}[h!]
\centering
\caption{Node potentials (in Volts) from experiments. $\mathbf{V^{(i)}}$ denotes data from the circuit in which node $i$ is grounded or the unperturbed circuit ($o$).  }
\begin{tabular}{l c c c c }
Experiment & Node 1 & Node 2 & Node 3 & Node 4 \\
\hline    
$\mathbf{V^{(1)}}$ & 0 & 0 & 0 & 0 \\
$\mathbf{V^{(2)}}$ & 4.41 & 0 & 0.85 & 0.16 \\
$\mathbf{V^{(3)}}$ & 5.37 & 2.54 & 0 & 0.12 \\
$\mathbf{V^{(4)}}$ & 7.2 & 5.39 & 4.25 & 0 \\
$\mathbf{V^{(o)}}$ & 7.55 & 5.95 & 4.95 & 1.23 \\
\end{tabular}
\label{ExampleData}
\end{table}

\subsection{Burned resistor detection}
\label{BurnedResistorDetection}
We have found that the experimental network technique can be taught to students using interesting, real-world applications. For example, students have been asked to use the approach to detect burned resistors. Imagine a circuit running an important electrical device, \eg life support equipment, which must remain running. This complicates any repairs if, for example, a resistor burns. The node potentials can be measured with a voltmeter, making it clear that the device is malfunctioning; however, it is not generally possible to pinpoint the burned resistor by examining potentials since most, if not all, will be modified due to the connectivity of the circuit. Further, an ohmmeter cannot be used to measure the resistance of the components while current is flowing through them. Fortunately, the network method can identify the burned resistor, which can then be replaced, without having to shut down the device.

This idea will, again, be demonstrated on the circuit of Fig. \ref{ExampleCircuit}. We have already determined $\mathbf{A}$ for the circuit and know the normal operating potentials, $\mathbf{V^{(o)}}=\left[7.55 \quad 5.95 \quad 4.95 \quad  1.23\right]$, Eq. \ref{ExampleA} and Table \ref{ExampleData}, respectively. If a component burns, the conductance will decrease (resisitance increases), resulting in a change in the elements of $\mathbf{A}$, due to Eq. (\ref{Coefs:a}), and a modification of the node potentials. For example, if $R_{34}$ burns, increasing its resistance to 50~M$\Omega$, the node potentials will be $\mathbf{V_\textrm{b}^{(o)}}=\left[9.03 \quad 8.25 \quad 8.45 \quad  0.49\right]$. It is not immediately obvious that $R_{34}$ was the component that suffered the damage. 

The network method can be implemented again, now for the malfunctioning circuit. It is non-invasive, only requiring the measurement of the new potentials for the normal and perturbed circuits. Previously, the perturbations were made by grounding each node individually, which was simply chosen for convenience. In this scenario, it might not be prudent to implement such a drastic change in the circuit while it is still in operation. A slight deviation from the normal node potentials can be used instead; the results will be unaffected. After completing the experiments and calculations for the burned circuit, the new $\mathbf{A}$ matrix is,
\begin{equation}
\label{BurnedA}
\mathbf{A_\mathrm{b}} =
\begin{pmatrix}
     1 & -0.43 & -0.14 & 0 \\
-0.47 & 1 & -0.47 & -0.06 \\
-0.25 & -0.75 & 1 & 0 \\
0 & -0.06 & 0 & 1 \\
\end{pmatrix}.
\end{equation}
Equations (\ref{ExampleA}) and (\ref{BurnedA}) can be compared. The elements $A_{\mathrm{b},34}$ and $A_{\mathrm{b},43}$ have now gone to zero, indicating that the conductance between nodes three and four has decreased significantly. It can be concluded that this component was damaged.

Such a comparison is not an efficient means of locating the burned component, especially if the circuit is large. A simple algorithm has been developed to easily identify the damaged resistor. It is based on the fact that not all of the elements of the matrix are modified when a component burns. In the example above, $A_{34}$ clearly changes since $A_{ij}=-G_{ij}/G_{ii}$ (by convention, $A_{ii}=1$). The drastic modification of $G_{34}$ is reflected in this element. $A_{43}$ is affected similarly. The coefficients of self-conductance for nodes three and four are also changed. They decrease due to the reduction of the conductivity. This has the effect of increasing all other elements in rows three and four, even if the associated resistors were unaffected by the burning. Finally, any elements not associated with the nodes three and four should remain unchanged. All of this can be observed by examining the difference between $\mathbf{A}$ and $\mathbf{A_b}$, \ie
\begin{equation}
\label{DeltaA}
\Delta \mathbf{A} = \left|\mathbf{A}\right| - \left|\mathbf{A_\mathrm{b}}\right| =
\begin{pmatrix}
     0 & 0 & 0 & 0 \\
0 & 0 & 0 & 0 \\
-0.07 & -0.20 & 0 & 0.27 \\
0 & -0.01 & 0.19 & 0 \\
\end{pmatrix}.
\end{equation}
The $\Delta A_{34}$ and $\Delta A_{43}$ elements are positive since their associated conductances decreased with the burning. The remaining elements in those rows are negative due to the decrease of the coefficients of self-conductance. All other elements have been unaffected by the burned component. Thus, the damaged component can be identified by searching for rows with non-zero elements. 

The concepts presented in this section have been verified on a variety of circuits, ranging from the simple to the quite complex. A circuit containing 18 nodes and 30 components cannot practically be analyzed by traditional methods, yet the network approach can be efficiently employed by students with common laboratory equipment. Unlike the straightforward implementation of Kirchhoff's laws, the network approach extracts the interactions between the nodes (conductances). This fact can be used to determine the resistors connecting nodes, and the equations describing the node potentials, for black box circuits in which the components are hidden. Furthermore, the approach automatically combines resistors in parallel and series since the elements of $\mathbf{A}$ are determined by the effective conductances between the nodes. 

The node method for a circuit of resistors produces a linear system of equations. The network approach, approximate for many networks, is exact for such systems due to its linear nature. Here, we have exploited this mathematical relationship to study circuits. Interestingly, before the advent of the handheld calculator, the relationship was used in reverse; circuits were used for algebraic calculations.\cite{Thesis} If nonlinear elements are present in the circuit, the network equations approximate the actual circuit behavior. In spite of the nonlinearity, the network model can accurately describe node potentials and can even be used to control the system, an important topic in network research.\cite{unpub}

\section{Conclusion}
We have demonstrated that the network modeling methodology can be used to obtain equations describing complex resistor circuits. These equations are the same as those obtained through more traditional means. The approach is unique in that it is experimental, rather than analytical, and well suited for complex arrays of resistors. The ease of implementation, to the extent that it resembles a recipe, is convenient; however, we do not suggest it to be a substitute for traditional instruction. It is simply an efficient means of obtaining the circuit equations when other techniques are not practical.

The connection between the circuit and network equations has, for the first time, provided physical interpretation of the elements of the $\mathbf{A}$ matrix. Each $A_{ij}$, $(i \neq j)$, represents the amount of influence the $ij$ channel has relative to all of the other connections to node $i$. This feature is not unique to circuits; it arises in systems where in the interactions between nodes is proportional to the difference in node parameter values.

We have found that this material is suitable for undergraduate students beyond the introductory courses, perhaps as an exercise in intermediate or advanced laboratories. It can be used as a springboard into discussions of networks, systems of equations, and the agreement between theory and experiment. Furthermore, it demonstrates the connection between research and the classroom.

\begin{acknowledgments}
The authors would like to thank Stephen Tsui for review of the manuscript as well as Jeniffer Allen and Jared Bland for discussions. This work was supported by the School of Natural Sciences and Mathematics and the Grants Office of The Richard Stockton College of New Jersey.
\end{acknowledgments}


\begin{thebibliography}{99}

\bibitem{AJPrandomwalk} Raymond A. Sorensen, ``The random walk method for dc circuit analysis,'' Am. J. Phys. \textbf{58}(11), 1056--1059 (1990).

\bibitem{AJPnode} David A. Giltinan, ``Multiloop dc circuits by source conversion and nodal analysis,'' Am. J. Phys. \textbf{62}(7), 645--647 (1994).

\bibitem{AJPvariational} D. A. Van Baak, ``Variational alternatives to Kirchhoff's loop theorem in dc circuits,'' Am. J. Phys. \textbf{67}(1), 36--44 (1999).

\bibitem{BookerText} Henry G. Booker, \textit{An Approach to Electrical Science} (McGraw Hill, New York 1959).

\bibitem{GunaratnePLOS} Gemunu H. Gunaratne, Preethi H. Gunaratne, Lars Seemann, and Andrei T\"{o}r\"{o}k, ``Using Effective Subnetworks to Predict Selected Properties of Gene Networks,'' PLoS ONE \textbf{5}(10), e13080 (2010).

\bibitem{ShulmanBiophysJ} Jason Shulman, Lars Seemann, and Gemunu H. Gunaratne, ``Effective Models of Periodically Driven Networks,'' Biophys. J. \textbf{101}(11), 2563--2571 (2011).


\bibitem{ShulmanBiophysRev} Jason Shulman, Lars Seemann, Gregg W. Roman, and Gemunu H. Gunaratne, ``Effective Models for Gene Networks and their Applications,'' Biophys. Rev. Lett. \textbf{7}, 41--70 (2012).

\bibitem{Gardner} Timothy S. Gardner, Diego di Bernardo, David Lorenz, and James J. Collins, ``Inferring Genetic Networks and Identifying Compound Mode of Action via Expression Profiling'', Science \textbf{301}, 102--105 (2003).

\bibitem{unpub} Jason Shulman, Alexander Mo, Killian Ryan, and Gemunu H. Gunaratne, unpublished.

\bibitem{Gardner2} Timothy S. Gardner and Jeremiah J. Faith, ``Reverse-engineering transcription control networks,'' Physics of Life Reviews \textbf{2}, 65--88 (2005).

\bibitem{NoteExact}It should be noted that condition 1 is sufficient to produce an exact model of the network. Condition 2 is necessary to produce an exact \textit{and complete} model. Without this condition, the effect of node three on nodes one and two would not be included. See Eqs. (\ref{MasterEx}) for an example.

\bibitem{SupplementaryMaterials} In the archive. Can contact for supplementary materials.

\bibitem{FoundationsText} Anant Agarwal and Jeffrey Lang, \textit{Foundations of Analog and Digital Electronic Circuits} (Morgan Kaufmann, San Francisco 2005).

\bibitem{NoteCoeff} The coefficient between the \textit{i}\textsuperscript{th} and \textit{j}\textsuperscript{th} nodes, $G_{ij}$, is often called the mutual conductance between the two nodes.

\bibitem{NoteCoeff2} This is rather obvious from Eq. (\ref{SelfCond}) and its generalization; however, to the best of the authors' knowledge, the most recent mention of this fact is from a textbook published in 1966, and reiteration seems appropriate. The citation for the textbook is Charles Close, \textit{The analysis of Linear Circuits} (Harcourt, Brace \& World, Inc., New York, 1966).

\bibitem{Simulation} Simulations were performed with 5spice software which can be found at \url{http://www.5spice.com/}.

\bibitem{Thesis} Donald Johnson, Masters Thesis, University of Wisconsin, 1953.


\end{thebibliography}
\end{document}